# Algebraic Determination of Spectral Characteristics of Rovibrational States of Diatomic Molecules. II. Analysis of the Vibration−Rotation Interaction by Means of the Factorization Method

S. A. Astashkevich[*]

St. Petersburg State University, Peterhof, St. Petersburg, 198904 Russia

**Abstract**

An algebraic model taking into account the influence of the molecular rotation on the wave functions of vibrational−rotational states of the diatomic molecule using the formalism of the ladder operators and an expansion in a small parameter $\varepsilon_e$ characterizing the vibrational−rotational interaction have been proposed for the potentials whose the creation and annihilation operators can be constructed. Expressions for the expansion of the wave function $\varphi_{v,J}$ of the states with the vibrational $v$ and rotational $J$ quantum numbers in a set of wave functions $\varphi_{v\pm l,0}$ of the rotationless vibrational states (with $0 \leq l \leq 2$) have been obtained of the second order in the parameter $\varepsilon_e$. Using these expressions and the formulas obtained in our previous paper the algebraic expressions for the expansion of the dependences of matrix elements $\langle v, J | f(r) | v', J' \rangle$ in a set of matrix elements on the wave functions of the rotationless ground vibrational states have been obtained for arbitrary functions of internuclear distance $f(r)$ and arbitrary values of $v$, $v'$ and $J$, $J'$. The cases of the harmonic oscillator and the Morse potential were analyzed. The explicit algebraic formulas for the vibrational dependences of the first five Herman−Wallis coefficients have been derived for arbitrary values of the vibrational quantum numbers of the combining states. The possibility of taking into account the influence of non−adiabatic effects on the vibrational and rotational dependences of matrix elements is analyzed.

---

[*] *E-mail address:* astashkevich@mail.ru

# 1. Introduction

In our previous paper [1] explicit algebraic expressions for the expansion of the vibrational matrix elements $\langle v|f(r)|v'\rangle$ in series of matrix elements on the wave functions of the ground vibrational state have been obtained for arbitrary values $v$ and $v'$, arbitrary sufficiently differentiable functions of the internuclear distance $f(r)$ and the potential curves whose ladder operators can be constructed, for example, for the harmonic oscillator and the Morse potential. The rotation of the molecule was disregarded in [1]. Present paper is devoted to further development of the algebraic approach proposed in [1] by taking into account the rotation of the molecule and the influence of the adiabatic effect of the vibrational–rotational interaction on the dependence of the vibrational matrix elements on the vibrational and rotational quantum numbers.

Information about the influence of the molecular rotation on various (radiative, electric, magnetic) characteristics of the diatomic molecule is of great interest as for study of the adiabatic (the vibrational–rotational interaction) and the non–adiabatic (the electronic–rotational interaction) effects on these characteristics [2–4] so important for many physical and chemical applications [4–6]. Obtaining data about these characteristics requires information about the dependencies of the vibrational matrix elements on the rotational quantum numbers of combining states of the molecule i.e. the taking into account the effect of vibrational–rotational interaction. For this study different methods are used: 1) the numerical methods that consist in the numerical solution of the vibrational Schrödinger equation [7], 2) semi–analytical methods based on various approximations (expansion in the Dunham series, the perturbation theory, etc.) and assuming (at least for the determination of non–energetic characteristics) numerical or algebraic calculations of the vibrational matrix elements with different values of $v$ and $v'$ [8–11]; 3) various algebraic methods (the Lie algebra [12], the method factorization [13], the using of special (hypergeometric) functions [14] and the theorems of quantum mechanics (hypervirial, Hellmann–Feynman, the sum rules) [13, 15–18]).

Semi–analytical methods in contrast to numerical methods allow partly conducting analytical studies of the influence of various effects (mechanical, electro–optical [11], etc.) on characteristics of the vibrational–rotational states of molecules. However these methods either require numerical calculations of matrix elements for the vibrational wave functions for different values of $v$ and $v'$ or give only some recurrence relations for required dependences of molecule characteristics on the vibrational quantum numbers of combining states, and obtained explicit formulas are limited to relatively small values of the differences quantum numbers *v* and *v'* of combining vibrational states.

Such a situation occurs, for example, in the case of determining the Herman−Wallis [4, 19]. Different techniques is used to obtain them: the perturbation theory for an arbitrary potential energy curve [9] and the potential curve in the form of the expansion in the Dunham parameter [8, 10], using of the hypervirial theorem and the sum rules [18], the formalism of polynomials on quantum numbers [11], and others. Explicit analytic expressions for these coefficients available in the literature are limited to values of the differences of the vibrational quantum numbers $\Delta v = |v - v'| \leq 7$. It should be noted that with increasing value $\Delta v$ one have to use the higher−order perturbation theory that leads, accordingly to the opinion of the authors of the works, to a "fairly cumbersome expressions" already for $\Delta v = 4$ [11] and "too long and extremely cumbersome" expressions for $\Delta v = 8$ [10] (see also [8]). Explicit analytical expressions for the dependence of the Herman−Wallis coefficients on the vibrational quantum numbers of combining states for arbitrary values of these quantum numbers is lacking in the literature so far. At the same time these expressions are actual for the determination of the semi−empirical dependence of the electronic characteristics of molecules on the internuclear distance (the dipole moment, the polarizability tensor coefficients, etc.) [4].

Therefore there is significant interest in the development of algebraic methods for determining the vibrational and rotational dependences of different characteristics of diatomic molecules that would provide the necessary mathematical foundation for more deep and systematic analysis of the influence of the different effects on these dependences on the vibrational and rotational quantum numbers. It should be noted that the analytical expression for the vibrational matrix elements on the rotational quantum numbers of combining states were obtained only for a few model functions (polynomial, exponential) for the Morse [13−15] and Kratzer (Kratzer) [16, 17] potentials. Also we emphasized that the literature contains no explicit (non−recurrence) algebraic relations for the vibrational matrix elements of an arbitrary function of internuclear distance on the rotational quantum numbers of combining states of the molecule. The purpose of present paper is to obtain such expressions using the formalism of creation and annihilation operators and formulas derived in the paper [1].

## 2. Basic notations

We consider the non−relativistic Hamiltonian of the diatomic molecule taking into account the rotation of the molecule and neglecting the non−adiabatic effects

$$\hat{H}(r) = -\frac{\hbar^2}{2\mu}\frac{d^2}{dr^2} + V(r) + \frac{\hbar^2}{2\mu r^2}J(J+1) ,\qquad(1)$$

here

$$V(r) = V_{BO}(r) + \Delta V_{ad}(r) ,$$

$V_{BO}(r)$ is the potential of the molecule in the Born−Oppenheimer approximation; $\Delta V_{ad}(r)$ is the adiabatic correction to the potential depending on the reduced mass of the molecule $\mu$; $r$ is the internuclear distance. Consider the case when the potential is expressed by an analytical dependence defined by the parameters $\vec{\xi}(J) = (\xi_1(J), \xi_2(J), ..., \xi_n(J))$. Expanding the term in the Hamiltonian associated with the rotation of the molecule (see Eq. (1)) in series in a velocity of the equilibrium internuclear distance $r_e$ in the parameter associated with a particular type of potential [14, 20, 21] the Hamiltonian of the molecule can be reduced up to a certain order accuracy in this parameter to the following form:

$$\hat{\tilde{H}}(r) = -\frac{\hbar^2}{2\mu}\frac{d^2}{dr^2} + U(\vec{\xi}(J), r) ,\qquad(2)$$

here $U(\vec{\xi}(J), r)$ is the potential that has the same form as the potential $U(\vec{\xi}(0), r)$ but depends on the parameters $\vec{\xi}(J)$. The Schrödinger equation for this Hamiltonian is

$$\hat{\tilde{H}}(r)\varphi_{v,J}(r) = E_{v,J}\varphi_{v,J}(r),$$

here $\varphi_{v,J}(r)$ and $E_{v,J}$ are the wave function and the energy of the vibrational−rotational states with quantum numbers $v$ and $J$ correspondingly.

The paper [1] was devoted to the algebraic analysis of the matrix elements on the wave functions of the rotationless vibrational states for the potentials possessing a dynamical symmetry whose the creation and annihilation operators can be constructed by means of the variable substituting $y = y(r)$. These operators have the following properties:

$$\hat{K}^+ \varphi_v(y) = \hat{a}_v^+ \varphi_{v+1}(y) ,  \qquad (3)$$

$$\hat{K}^- \varphi_v(y) = \hat{a}_v^- \varphi_{v-1}(y) . \qquad (4)$$

We consider a fairly general form of the creation and annihilation operators

$$\hat{K}^\pm = \frac{\partial}{\partial y} \hat{b}_v^\pm + \hat{c}_v^\pm(y) . \qquad (5)$$

The operator coefficients $\hat{a}_v^\pm$, $\hat{b}_v^\pm$ and functions $\hat{c}_v^\pm(y)$ depend on the value of the vibrational quantum number and the parameters of the molecule $\vec{\xi}(0) = (\xi_1(0), \xi_2(0), ..., \xi_n(0))$ (see the paper [1]). The partial derivative in formula (5) is taken with respect to $y$ for the fixed values of these parameters.

The objective of present work is to obtain formulas for the matrix elements of the wave functions of the vibrational–rotational states in terms of the matrix elements of the wave functions of pure vibrational states (also called pure vibrational states). The solution of this problem we conduct into two stages: first, we express the wave functions of the vibrational–rotational state in terms of the wave functions of the pure vibrational states and, second, we obtained according expressions for matrix elements on these wave functions.

### 3. Wave functions of vibrational–rotational states

Now we express the wave function $\varphi_{v,J}(y)$ in terms of wave functions of pure vibrational states. For it we expand this wave function in a Taylor series with respect to parameters $\Delta\vec{\xi}(J) = \vec{\xi}(J) - \vec{\xi}(0)$ that is equal to the difference of the parameters of the potential curve of rotating and rotationless molecules in the vicinity of the parameters of rotationless molecule $\vec{\xi}(0)$. Taking into account the dependence of the dynamical variable $y$ parameters on the parameters $\vec{\xi}(0)$ we obtain the following expression:

$$\varphi_{v,J}(y) = \varphi_v(y) + \sum_{i=1}^n \left[ \frac{\partial y}{\partial \xi_i} \frac{\partial}{\partial y} \varphi_v(y) \right] \Delta\xi_i(J) + \\ + \frac{1}{2} \sum_{i=1}^n \sum_{k=1}^n \left[ \frac{\partial^2 y}{\partial \xi_i \partial \xi_k} \frac{\partial}{\partial y} \varphi_v(y) + \frac{\partial y}{\partial \xi_i} \frac{\partial y}{\partial \xi_k} \frac{\partial^2}{\partial y^2} \varphi_v(y) \right] \Delta\xi_i(J) \Delta\xi_k(J) + ... \qquad (6)$$

Derivatives in this formula should be taken in the parameters $\vec{\xi}(0)$. Formula (6) can be written as an expansion of the derivatives $\partial^l \varphi_v(y)/\partial y^l$:

$$\varphi_{v,J}(y) = \varphi_v(y) + A^{(1)}_{v,J}(y)\frac{\partial}{\partial y}\varphi_v(y) + A^{(2)}_{v,J}(y)\frac{\partial^2}{\partial y^2}\varphi_v(y) + ... \qquad (7)$$

We assume that the relative change in the parameters $\vec{\xi}(J)$ associated with the rotation of the molecule is small: $\left|\Delta\xi_i(J)/\xi_i(0)\right| \ll 1$. Then the functions before the derivatives in Eq. (7) in the second order in the parameters $\Delta\xi_i(J)$ are

$$A^{(1)}_{v,J}(y) = \sum_{i=1}^{n}\frac{\partial y}{\partial \xi_i}\Delta\xi_i(J) + \frac{1}{2}\sum_{i=1}^{n}\sum_{k=1}^{n}\frac{\partial^2 y}{\partial \xi_i \partial \xi_k}\Delta\xi_i(J)\Delta\xi_k(J), \qquad (8)$$

$$A^{(2)}_{v,J}(y) = \frac{1}{2}\sum_{i=1}^{n}\sum_{k=1}^{n}\frac{\partial y}{\partial \xi_i}\frac{\partial y}{\partial \xi_k}\Delta\xi_i(J)\Delta\xi_k(J). \qquad (9)$$

The changing of the parameters $\xi_i(J)$ associated with the molecular rotation can be represented as an expansion in small parameter $\varepsilon_e$ that characterizes the vibrational–rotational interaction and determined by the parameters $\vec{\xi}(0)$:

$$\Delta\xi_i(J) = \varepsilon_e \xi_i^{(1)} J(J+1) + (\varepsilon_e)^2 \xi_i^{(2)} J^2(J+1)^2 + ... \qquad (10)$$

In present paper as the expansion the Dunham approximation [20] for the harmonic oscillator and Pekeris approximation [21] for the Morse potential were used. For the harmonic oscillator the parameter $\varepsilon_e = 4(B_e/\omega_e)^2$ have been taken and for the Morse potential the parameter $\varepsilon_e = B_e/D_e$ have been taken ($B_e$ and $\omega_e$ – the rotational and vibrational constants, $D_e$ – the dissociation energy of the Morse potential). The parameters $\omega_e$ and $\alpha$ are related by the formula $\alpha = (\mu\omega_e)^{1/2}/\hbar$. As the parameters $\xi_i(0)$ ($i$ = 1,2) the parameters ($r_e$, $\alpha$) (the harmonic oscillator) and ($r_e$, $q$) (the Morse potential) have been taken (see Table).

**Table.** The formulas for the dependence of the dynamic variable on the internuclear distance $y=y(r)$, the functions $h_1(y)$ and $h_2(y)$ and the coefficients $a_v^{\pm}$ и $d_v^l$ ($l = -1, 0, 1$) for the harmonic potential and the Morse potential. The following designations are used: $b = \alpha r_e$ and $q = (8\mu D_e)^{1/2}/\alpha \hbar$.

| Parameters | Potentials | |
|---|---|---|
| | Harmonic oscillator | Morse potential |
| $U(\vec{\xi},r)$ | $\dfrac{\hbar^2}{2\mu}\alpha^4(r-r_e)^2$ | $D_e[\exp(-2\alpha(r-r_e)) - 2\exp(-\alpha(r-r_e))]$ |
| $y=y(r)$ | $\alpha(r-r_e)$ | $q\exp(-\alpha(r-r_e))$ |
| $h_1(y)$ | $\dfrac{3}{4}y - b$ | $\dfrac{1}{2b}\left(\dfrac{3}{b}-1\right)y$ |
| $h_2(y)$ | $-\dfrac{27}{32}y + \dfrac{9}{4}b$ | $\dfrac{1}{b^2}\left(1-\dfrac{3}{2b}\right)y$ |
| $a_v^{-}$ | $\sqrt{v}$ | $\sqrt{v(q-v)}$ |
| $a_v^{+}$ | $\sqrt{v+1}$ | $\sqrt{(v+1)(q-v-1)}$ |
| $d_v^{-1}$ | $\dfrac{1}{\sqrt{2}}$ | $-\dfrac{1}{2(q-2v)}\sqrt{\dfrac{q-2v-1}{q-2v+1}}$ |
| $d_v^{0}$ | $0$ | $\dfrac{q}{2(q-2v)(q-2v-2)}$ |
| $d_v^{1}$ | $-\dfrac{1}{\sqrt{2}}$ | $\dfrac{1}{2(q-2v-2)}\sqrt{\dfrac{q-2v-1}{q-2v-3}}$ |

Using Eq. (10) formulas (8) and (9) in the second order of the parameter $\varepsilon_e$ can be written as

$$A_{v,J}^{(1)}(y) = h_1(y)\varepsilon_e J(J+1) + h_2(y)(\varepsilon_e)^2 J^2(J+1)^2 , \qquad (11)$$

$$A_{v,J}^{(2)}(y) = h_3(y)(\varepsilon_e)^2 J^2(J+1)^2 , \qquad (12)$$

where the functions $h_1(y)$, $h_2(y)$ and $h_3(y)$ are

$$h_1(y) = \sum_{i=1}^{n} \frac{\partial y}{\partial \xi_i} \xi_i^{(1)} , \qquad (13)$$

$$h_2(y) = \sum_{i=1}^{n} \frac{\partial y}{\partial \xi_i} \xi_i^{(2)} + \frac{1}{2}\sum_{i=1}^{n}\sum_{k=1}^{n} \frac{\partial^2 y}{\partial \xi_i \partial \xi_k} \xi_i^{(1)} \xi_k^{(1)} , \qquad (14)$$

$$h_3(y) = \frac{1}{2}(h_1(y))^2 . \qquad (15)$$

The formulas for the functions $h_1(y)$ и $h_2(y)$ for the harmonic oscillator and the Morse potential is given in Table. It is seen that these functions have linear dependences on *y*.

We express the differential operator in Eq. (7) with regard to the creation and annihilations operators. One can see that formula (5) gives a system of two linear equations for the creation and annihilation operators. Knowing the explicit form of these operators [22, 23] the expression for the operator $\partial/\partial y$ is

$$\frac{\partial}{\partial y} = \hat{K}^+ \hat{d}_v^1 + \hat{K}^- \hat{d}_v^{-1} + \hat{d}_v^0 . \qquad (16)$$

Formulas for the coefficients $d_v^l$ ($l = -1, 0, 1$) for the harmonic oscillator and the Morse potential obtained from the data given in [23] are in Table. Substituting the expression for the differentiation operator (Eq. (16)) into Eq. (7) and using Eqs. (3) and (4) we obtain

$$\varphi_{v,J}(y) = \sum_{l=-2}^{2} B_{v,J}^{(l)}(y)\varphi_{v+l}(y) . \qquad (17)$$

With the help of Eqs. (7), (11)–(16) the expression for the coefficients $B_{v,J}^{(l)}(y)$ is

$$B_{v,J}^{(l)}(y) = \sum_{p=0}^{2}\sum_{q=1}^{3} \eta_{p,q}^{(l)}(v)(h_q(y))^{1-\delta_{p,0}}(\varepsilon_e)^p J^p(J+1)^p \ , \qquad (18)$$

here $\delta_{p,0}$ is the Kronecker symbol. The expressions for the non–zero coefficients $\eta_{p,q}^{(l)}$ for different values of $l$, $p$ and $q$ ($-2 \leq l \leq 2$, $0 \leq p \leq 2$ and $1 \leq q \leq 3$) are

$$\eta_{0,1}^{(0)}(v) = 1$$

$$\eta_{1,1}^{(0)}(v) = \eta_{2,2}^{(0)}(v) = d_v^0$$

$$\eta_{2,3}^{(0)}(v) = a_{v+1}^- a_v^+ d_{v+1}^{-1} d_v^1 + a_{v-1}^+ a_v^- d_{v-1}^1 d_v^{-1} + \left(d_v^0\right)^2 \qquad (19)$$

$$\eta_{1,1}^{(\pm 1)}(v) = \eta_{2,2}^{(\pm 1)}(v) = a_v^\pm d_v^{\pm 1}$$

$$\eta_{2,3}^{(\pm 1)}(v) = a_v^\pm d_v^{\pm 1}\left(d_{v\pm 1}^0 + d_v^0\right)$$

$$\eta_{2,3}^{(\pm 2)}(v) = a_{v\pm 1}^\pm a_v^\pm d_{v\pm 1}^{\pm 1} d_v^{\pm 1}$$

The coefficients $\eta_{p,q}^{(l)}$ are equal to zero for all other values of the indices $l$, $p$ and $q$. One should be noted that Eqs. (19) contain not the operators $\hat{a}_v^\pm$ и $\hat{d}_v^l$ but the numerical coefficients $a_v^\pm$ and $d_v^l$ corresponding the action of these operators on the wave functions (see the paper [1]). An analysis of Eqs. (18) and (19) shows that the coefficients $B_{v,J}^{(0)}(y)$, $B_{v,J}^{(\pm 1)}(y)$ and $B_{v,J}^{(\pm 2)}(y)$ have values of the order 1, $\varepsilon_e J(J+1)$ and $(\varepsilon_e)^2 J^2(J+1)^2$ correspondingly.

It can seen that the wave function of the vibrational–rotational level can be expressed as a sum of the five wave functions of pure vibrational states with the values of the vibrational quantum number $v$, $v\pm 1$ and $v\pm 2$ (see Eq. (17)) of the second order accuracy in the parameter $\varepsilon_e$. Note that the scheme of our analysis can be also applied to the obtaining of the corresponding expression for the $\varphi_{v,J}(y)$ up to higher order accuracy in the parameter $\varepsilon_e$. One can show that the wave function of the vibrational–rotational state can be expressed as a sum of ($l_{max}$+1) the wave functions of pure vibrational states with the values of the vibrational quantum number ($l$ = 0, 1, ..., $l_{max}$) of the $l_{max}$–order accuracy in the parameter $\varepsilon_e$.

To determine the coefficients $B_{v,J}^{(l)}(y)$ (Eq. (17)) it is required the information about: 1) the dependences of the parameters $\vec{\xi}(J)$ of the potential $U(\vec{\xi}(J),r)$ on the rotational quantum number (Eq. (10)), 2) the dependences of the dynamical variable $y$ on the parameters $\vec{\xi}(J)$ (Eqs. (13)–(15) and (18)), 3) the coefficients $a_v^{\pm}$ (Eqs. (3), (4) and (19)) and the coefficients $d_v^l$ (Eqs. (16) and (19)) defined by the explicit form of the creation and annihilation operators for a specific type of the potential curve.

## 4. Matrix elements

We obtain an algebraic expression for the matrix elements $\langle v, J | f(r) | v+m, J+M \rangle$ of an arbitrary function of the internuclear distance on the wave functions with the values of the vibrational and rotational quantum numbers of combining states $v$, $J$ and $v'$ = $v + m$, $J'$=$J$+$M$ where $-v \leq m \leq v_{max}-v$ and $-J \leq M$ ($v_{max}$ is equal to the maximum value of the vibrational quantum number for the potentials having the dissociation threshold, and $v_{max}$=∞ for potentials that do not have such a threshold, for example, for the harmonic oscillator). For this we represent the wave functions of the combined vibrational–rotational states as an expansion of the wave functions of pure vibrational states (see Eq. (17)), and then substitute these formulas into the required expression for the matrix elements. Finally we obtain

$$\langle v, J | f(r) | v+m, J+M \rangle = \langle v, 0 | f(r) | v+m, 0 \rangle +$$
$$+ \sum_{p=0}^{2} \sum_{p'=0}^{2} \sum_{q=1}^{3} \sum_{q'=1}^{3} \sum_{l=-2}^{2} \sum_{l'=-2}^{2} \Big\{ (1-\delta_{p,0}\delta_{p',0}) \eta_{p,q}^{(l)}(v) \eta_{p',q'}^{(l')}(v+m) J^p (J+1)^p (J+M)^{p'} (J+M+1)^{p'} \quad (20)$$
$$\times (\varepsilon_e)^{p+p'} \left\langle v+l, 0 \Big| (h_q(y))^{1-\delta_{p,0}} (h_{q'}(y))^{1-\delta_{p',0}} f(y) \Big| v+m+l', 0 \right\rangle \Big\}$$

The integration of the wave functions in the matrix elements under the signs of summation in Eq. (20) is performed on the variable $y(\vec{\xi}(0))$ depending on the parameters of the potential of the rotationless molecule. The matrix elements to the right in Eq. (20) are taken on the wave functions of pure vibrational states. The formula for these matrix elements was obtained in the paper [1]. With the help of Eqs. (8)–(10), (14) and (15) from [1] we obtain:

$$\langle v,J|f(r)|v+m,J+M\rangle = \sum_{p=0}^{2}\sum_{p'=0}^{2}\sum_{q=1}^{3}\sum_{q'=1}^{3}\sum_{l=-2}^{2}\sum_{l'=-2}^{2}\sum_{j=0}^{v_{m,l,l'}}\left\{\chi_{p,p',q,q',j}^{(l,l')}(v,m)g_{p,p'}(J,M)\times\right.$$
$$\left.\times(\varepsilon_e)^{p+p'}\left\langle 0,0\left|\frac{d^{2j+|m+l'-l|}}{dy^{2j+|m+l'-l|}}\left[(h_q(y))^{1-\delta_{p,0}}(h_{q'}(y))^{1-\delta_{p',0}}f(y)\right]\right|0,0\right\rangle\right\}, \quad (21)$$

Here $v_{m,l,l'} = \min(v+l, v+m+l')$ and the coefficients in the formula (21) are as follows

$$g_{p,p'}(J,M) = J^p(J+1)^p(J+M)^{p'}(J+M+1)^{p'}, \quad (22)$$

$$\chi_{p,p',q,q',j}^{(l,l')}(v,m) = \gamma_j(v_{m,l,l'},|m+l-l'|)\eta_{p,q}^{(l)}(v)\eta_{p',q'}^{(l')}(v+m). \quad (23)$$

According to Eqs. (8)–(10) and (14) from [1] the coefficients $\gamma_j(v,|m|)$ are formulated as

$$\gamma_j(v,|m|) = (b_0^-)^{j+|m|}\frac{\prod_{k=j+1}^{v}a_{k+|m|}^-}{\prod_{k=1}^{v}a_k^-\prod_{k=1}^{j+|m|}a_k^-}\left(\sum_{i_1=j}^{v}b_{i_1+|m|}^-\sum_{i_2=j}^{i_1}b_{i_2+|m|}^-\sum_{i_3=j}^{i_2}b_{i_3+|m|}^-\cdots\sum_{i_j=j}^{i_{j-1}}b_{i_j+|m|}^-\right). \quad (24)$$

The matrix elements to the right in Eq. (21) are taken on the wave functions of the ground purely vibrational state. One can see that the dependence of matrix elements $\langle v,J|f(r)|v+m,J+M\rangle$ on the rotational quantum numbers of combining states is determined only by the factors $g_{p,p'}(J,M)$ and the dependence of these matrix elements on the vibrational quantum numbers is determined only by the factors $\chi_{p,p',q,q',j}^{(l,l')}(v,m)$.

Eq. (21) can be written in a general form as

$$\langle v,J|f(r)|v+m,J+M\rangle = \sum_{i=0}^{2}\sum_{\substack{k=0\\(i+k)\leq 2}}^{2} Z_{ik}(v,m)(\varepsilon_e)^{i+k}J^i(J+1)^i(J+M)^k(J+M+1)^k, \quad (25)$$

The summarization in Eq. (25) is limited the condition $(i+k)\leq 2$ because Eqs. (21) and (25) are obtained only up to second order accuracy in the parameter $\varepsilon_e$. The coefficients $Z_{ik}(v,m)$ are independent on the rotational quantum numbers of combining states and described by the following formula

$$Z_{ik}(v,m) = \sum_{l=-2}^{2}\sum_{l'=-2}^{2}\sum_{q=1}^{3}\sum_{q'=1}^{3}\sum_{j=0}^{v_{m,l,l'}} \left\{\chi_{i,k,q,q',j}^{(l,l')}(v,m)\times \right.$$
$$\left.\times\left\langle 0,0\left|\frac{d^{2j+|m+l'-l|}}{dy^{2j+|m+l'-l|}}\left[(h_q(y))^{1-\delta_{i,0}}(h_{q'}(y))^{1-\delta_{k,0}}f(y)\right]\right|0,0\right\rangle\right\}. \quad (26)$$

The main term in formula (26) corresponds to the matrix element on the wave functions of the rotationless molecule (see [1]):

$$Z_{00}(v,m) = \langle v,0|f(r)|v+m,0\rangle = \sum_{j=0}^{v_m}\gamma_j(v,|m|)\left\langle 0,0\left|\frac{d^{2j+|m|}}{dy^{2j+|m|}}f(y)\right|0,0\right\rangle,$$

here $v_m = \min(v,v+m)$. To determine the coefficients $Z_{ik}(v,m)$ (see Eq. (26)) it is sufficient to have the information about: 1) the wave function $\varphi_0(y)$ of the ground pure vibrational state 2) the functions $h_1(y)$ and $h_2(y)$ (Eqs. (13)–(15)) defined by the parameters of the potential curve and the dependencies of these parameters on the dynamical variable $y$; 3) the coefficients $a_k^-$ и $b_k^-$ (see Eqs. (3), (4), (19), (21), (23), and (24)) and the coefficients $d_v^l$ ($l = -1, 0, 1$) (see Eqs. (16) (19) (21), and (23)) determined by the explicit form of the creation and annihilation operators. All these data for the harmonic oscillator and the Morse potential are given in Tables of present paper and the paper [1].

It should be noted that to determine the matrix elements $\langle v,J|f(r)|v+m,J+M\rangle$ up to first order accuracy in the parameter $\varepsilon_e$ it is sufficient to know the matrix elements $\left\langle 0,0 \left| \frac{d^p}{dy^p} f_k(y) \right| 0,0 \right\rangle$ for the derivatives only two functions $f_0(y) = f(y)$ and $f_1(y) = h_1(y) f(y)$. The corresponding second order accuracy formulas need additionally the knowledge of the matrix elements of derivations yet two function $f_2(y) = h_2(y) f(y)$ and $f_3(y) = (h_1(y))^2 f(y)$ (see Eqs. (15), (25) and (26)). For the harmonic oscillator and the Morse potential functions there are only three such independent functions: $f_\kappa(y) = y^k f(y)$ ($0 \leq k \leq 2$) (see Table). The derivatives of these functions are taken for the following values of $p$: $p_{min} \leq p \leq (2v + m + 4)$ where $p_{min} = 0$ in the case $|m| \leq 4$ and $|m| - 4$ in the case $|m| > 5$.

An interesting consequence of the derived expressions is found: if the relation $f(y) = (h_1(y))^{-1}$ takes place then the factors $Z_{01}(v,m)$ and $Z_{10}(v,m)$ are equal to zero and the influence of the molecular rotation on the matrix elements manifests only in second order in the parameter $\varepsilon_e$ (see Eq. (25) and (26)).

## 5. Herman–Wallis factors

Using the developed approach we can obtain the expressions for the Herman–Wallis factors [4] determined as

$$F_{v,J}^{v',J'} = \left| \frac{\langle v, J | f(r) | v', J' \rangle}{\langle v, 0 | f(r) | v', 0 \rangle} \right|^2 . \tag{27}$$

Beginning with the work [19] these factors are usually represent as

$$F_{v,J}^{v',J'} = \sum_{k=0}^{k=K} C_k(v, v') \tilde{m}^k , \tag{28}$$

where the expression for the parameter $\tilde{m}$ is determined by the type of branch. The Herman–Wallis coefficients $C_k(v, v')$ do not depend on the rotational quantum numbers and are determined only by the specific form of the function $f(r)$, the type of the potential curve and the vibrational quantum numbers of combining states. Explicit expressions for the vibrational dependences of these coefficients for the potential curves whose ladder operators can be constructed can be derive from the formulas obtained higher. We obtain these expressions up to second order in the parameter $\varepsilon_e$. One consider two cases $J = J'$ and $J \neq J'$ separately.

i) The case $J = J'$. Using Eq. (25) and setting $\tilde{m} = J(J+1)$ we obtain the following formulas for the Herman–Wallis coefficients:

$$C_0(v,v') = 1 \quad, \tag{29}$$

$$C_1(v,v') = 2\varepsilon_e \tilde{Z}_1^+(v,v') \quad, \tag{30}$$

$$C_2(v,v') = (\varepsilon_e)^2 \left[ (\tilde{Z}_1^+(v,v'))^2 + \tilde{Z}_2^+(v,v') \right] \quad. \tag{31}$$

here and further the following notation are used

$$\tilde{Z}_1^\pm(v,v') = \tilde{Z}_{01}(v,v') \pm \tilde{Z}_{10}(v,v') \quad,$$

$$\tilde{Z}_2^\pm(v,v') = 2(\tilde{Z}_{02}(v,v') + \tilde{Z}_{20}(v,v') \pm \tilde{Z}_{11}(v,v')) \quad,$$

where

$$\tilde{Z}_{ik}(v,v') = \frac{Z_{ik}(v,v'-v)}{Z_{00}(v,v'-v)} \quad. \tag{32}$$

The coefficients $C_3(v,v')$ and $C_4(v,v')$ are of the order $(\varepsilon_e)^3$ and $(\varepsilon_e)^4$ correspondingly.

ii) The case $J \neq J'$. Denoting $\tilde{m} = [J'(J'+1) - J(J+1)]/2$ and $M = J'-J$ and using Eq. (25) we obtain

$$C_0(v,v') = 1 + \frac{M^2-1}{2}\varepsilon_e \tilde{Z}_1^+(v,v') + \left(\frac{M^2-1}{4}\right)^2 (\varepsilon_e)^2 \left[ \left(\tilde{Z}_1^+(v,v')\right)^2 + \tilde{Z}_2^+(v,v') \right], \tag{33}$$

$$C_1(v,v') = 2\varepsilon_e \tilde{Z}_1^-(v,v') + \frac{M^2-1}{2}(\varepsilon_e)^2 \left[ \tilde{Z}_1^+(v,v')\tilde{Z}_1^-(v,v') + \tilde{Z}_3(v,v') \right], \tag{34}$$

$$C_2(v,v') = \frac{2\varepsilon_e}{M^2}\tilde{Z}_1^+(v,v') + (\varepsilon_e)^2 \left\{ (\tilde{Z}_1^-(v,v'))^2 + \tilde{Z}_2^-(v,v') + \frac{M^2-1}{2M^2}\left[(\tilde{Z}_1^+(v,v'))^2 + \tilde{Z}_2^+(v,v')\right] \right\}, \tag{35}$$

$$C_3(v,v') = \frac{2(\varepsilon_e)^2}{M^2}\left[\tilde{Z}_1^+(v,v')\tilde{Z}_1^-(v,v') + \tilde{Z}_3(v,v')\right], \tag{36}$$

$$C_4(v,v') = \frac{(\varepsilon_e)^2}{M^4}\left[(\tilde{Z}_1^+(v,v'))^2 + \tilde{Z}_2^+(v,v')\right], \tag{37}$$

here

$$\tilde{Z}_3(v,v') = 2(\tilde{Z}_{02}(v,v') - \tilde{Z}_{20}(v,v')).$$

One can see that the coefficient $C_0(v,v')$ is of the order 1, the coefficients $C_1(v,v')$ and $C_2(v,v')$ are of the order of $\varepsilon_e$, and the coefficients $C_3(v,v')$ and $C_4(v,v')$ are of the order of $(\varepsilon_e)^2$ (see Eqs. (33)–(37)).

Thus, in order to calculate the Herman–Wallis coefficients up to second order accuracy in the parameter characterizing the vibrational–rotational interaction it is sufficient to know the values of this parameter (see Eq. (10)) and the coefficients $Z_{ik}(v,v'-v)$ (see Eqs. (25) and (32)) for the indices $0 \leq i, k \leq 2$ satisfying the condition $(i + k) \leq 2$.

It is important to note that in contrast to previous studies we obtained formulas for the explicit dependencies of the Herman–Wallis coefficients on the vibrational quantum numbers of combining states of the molecule for the all range of these quantum numbers (Eqs. (30)–(37)). These expressions can be used to determine the vibrational dependence of the Herman–Wallis coefficients for both the dipole transitions in the emission/absorption spectra and the Raman spectra. At this the substitutions $\tilde{m} = J+1$ and $\tilde{m} = -J$ should be made in Eq. (28) for the lines of the R- and P-branches correspondingly. For the lines of the S- and O-branches of the Raman spectra and electric quadrupole (or magnetic) transitions the substitutions $\tilde{m} = 2J+3$ and $\tilde{m} = -2J+1$ shiold to made in Eq. (28) correspondingly. As a function $f(r)$ in Eq. (27) the dependence of the appropriate transition moment on the internuclear distance should be substituted in this case of the IR emission/absorption spectra and the dependences of the isotropic and anisotropic components of the polarizability tensor should be substituted for the Raman spectra [24].

## 6. Taking into account of non−adiabatic effects

So far we neglected the influence of non−adiabatic effects. In the more general case that takes into account both adiabatic and non−adiabatic effects of intermolecular interactions the Hamiltonian of the molecule can be written as [25, 26]

$$\hat{H}(r) = -\left[1+\tilde{\beta}(r)\right]\frac{\hbar^2}{2\mu}\frac{d^2}{dr^2} + \left[1+\tilde{\alpha}(r)\right]\frac{\hbar^2}{2\mu r^2}J(J+1) + V(r), \qquad (38)$$

here $\tilde{\alpha}(r)$ and $\tilde{\beta}(r)$ – the functions describing non−adiabatic effects associated with the rotation and the vibration of the molecule correspondingly (these functions describe the interaction of the electronic state under investigation with other electronic states). For the most cases of practical interest the $\tilde{\beta}(r)$ can be approximated by a constant $\tilde{\beta}(r) \approx \tilde{\beta}_0$ according to [26]. Then the Hamiltonian can be reduced to the "normal" form assuming that the factor before the operator $d^2/dr^2$ in Eq. (38) is equal to $\hbar^2/2\tilde{\mu}$ where $\tilde{\mu} = \mu/(1+\tilde{\beta}_0)$.

Setting the function $V(r)$ in the form of the analytical potential $V(r) = U(\vec{\xi}(J), r)$ and expanding this potential and also the dependences $\tilde{\alpha}(r)$ and $1/r^2$ in a series of a small parameter in the velocity of $r_e$ the Hamiltonian of the molecule can be reduced to the form described by Eq. (2) of some accuracy order in the parameter. At this the substitutions $\mu$ on $\tilde{\mu}$ and $\vec{\xi}(J)$ on $\vec{\tilde{\xi}}(J)$ should be made in Eq. (2). Thus, the formulas obtained higher in the present paper can be used even for the case when non−adiabatic effects take place.

For highly exited electronic states of molecules in the case when the function $\tilde{\alpha}(r)$ is determined by the electronic−rotational interaction the matrix element on the electronic wave functions characterizing this interaction is independent of $r$ and equal to a known constant that depends on the symmetry of interacting electronic states (the "pure precession" approximation) [2]. In this case we can neglect the dependence $\tilde{\alpha}$ on the internuclear distance and set $\tilde{\alpha} = \tilde{\alpha}_0$. If also we suggest that the adiabatic correction to the Born–Oppenheimer potential curve can be approximated by the dependence[1] $\Delta V_{ad}(r) = A\hbar^2/(2\mu r^2)$ than the expression for the non−adiabatic Hamiltonian (Eq. (38)) can be reduced to formula (2) for the Hamiltonian without regard of non−adiabatic effects. To determine the matrix elements in this case Eqs. (25) and (26) obtained for the unperturbed case can be used. It should be substituted $\mu$ on $\tilde{\mu}$ and $J(J+1)$ on $J(J+1)\left[(1+\tilde{\alpha}_0)/(1+\tilde{\beta}_0)\right] + A/(1+\tilde{\beta}_0)$ in Eq. (2). Of course this approximation can be too rough at the study of spectral characteristics of real molecules and more complicated model can be required.

## 5. Conclusion

As a result of an algebraic analysis we obtain the expressions that allow us to express the matrix elements of the wave functions of the vibrational−rotational states as the set of matrix elements on the wave functions of the ground pure vibrational states ($v=0$, $J=0$). The deriving these expressions is based on: 1) the reduction of the potential of the rotating molecule to rotationless one by means of expanding the parameters of this potential in a small parameter in the vicinity of the equilibrium internuclear distance; 2) the knowledge of the explicit form of the creation and annihilation operators for the potential curve; 3) the expansion of the wave function of the vibrational−rotational state in series of a parameter characterizing the vibrational−rotational interaction.

Obtained expressions can be used for systematical algebraic analysis of the dependences of various characteristics of the molecules on the vibrational and rotational quantum numbers of combining states: the emission (or absorption) transitions probability (for dipole, quadrupole and higher order moments) of the electric (or magnetic) transitions in the IR spectra, the transition probabilities in the Raman spectra and other radiative (lifetimes, branching ratios), electrical (dipole moments) and magnetic (*g*-factors) characteristics of the

---

[1] Note that Eq. (38) (see [25]) is correct only for the $^1\Sigma$ state. Assuming this approximation for the dependence $\Delta V_{ad}(r)$ the adiabatic effects can be taking into account also for the electronic states with non-zero projection of the total electronic orbital angular momentum on the internuclear line ( $\Lambda \neq 0$). For the singlet states the relation $A = -2\Lambda^2$ takes place (see [2,3]).

vibrational−rotational and rovibrational states of diatomic molecules as well as various characteristics of the spectra of polyatomic molecules, for example, using the local modes model [27]. These expressions can be used not only for the singlet states but also for the multiplet electronic states provided that the influence of the effects associated with the electronic and nuclear spins on matrix elements is much less the influence of the vibrational−rotational interaction.